\documentclass[twocolumn, preprintnumbers,amsmath,amssymb]{revtex4}
\usepackage{cases}
\usepackage{amsmath}
\usepackage{amssymb}
\usepackage{amsfonts}
\usepackage{amssymb}
\usepackage{dcolumn}
\usepackage{bm}
\usepackage{epsfig}
\usepackage{bbm}
\usepackage{graphicx}
\usepackage{xcolor}
\usepackage{array}
\usepackage{subfigure}
\usepackage{hyperref}
\usepackage{mathrsfs}

\newcommand{\be}{\begin{equation}}
\newcommand{\ee}{\end{equation}}
\newcommand{\ba}{\begin{eqnarray}}
\newcommand{\ea}{\end{eqnarray}}

\newcommand{\gsim}{\mathrel{\hbox{\rlap{\lower.55ex \hbox {$\sim$}}
			\kern-.3em \raise.4ex \hbox{$>$}}}}
\newcommand{\lsim}{\mathrel{\hbox{\rlap{\lower.55ex \hbox {$\sim$}}
			\kern-.3em \raise.4ex \hbox{$<$}}}}

\hypersetup{colorlinks=true,
	breaklinks=true,
	pdfstartview=Fit,
	linkcolor=blue,
	citecolor=blue,
	urlcolor=blue}

\begin{document}

\title{The Realistic Scattering of Puffy Dark Matter}

\author{Wenyu Wang}
\email{wywang@bjut.edu.cn}
\author{Wu-Long Xu}
\email{wlxu@emails.bjut.edu.cn}
\affiliation{Faculty of Science, Beijing University of Technology, Beijing, China}
\author{Bin Zhu}
\email{zhubin@mail.nankai.edu.cn}
\affiliation{Department of Physics, Yantai University, Yantai 264005, China}

\begin{abstract}
If dark matter has a finite size, the intrinsic interaction responsible for the structure formation is inevitable from the perspective of dark matter self-scattering.  To describe the circumstance in which the binding force realizes the finite size dark protons, we first use the Eikonal approximation to simplify the convoluted scattering between dark protons into the case at the $t=0$ limit. The Chou-Yang model is then introduced to reduce the number of input parameters to one based on the simplicity and analyticity principle.
A new definition of velocity dependence and the corresponding implications on the small cosmological structures from Chou-Yang dark protons are shown clearly. Even though the parameter space is not fully covered, the numerical findings show that the amplitude coefficient can alter the self-scattering cross-section, allowing us to recover the excluded parameter space without using binding force. Finally, we demonstrate that the correct relic density from thermal freeze-out production prefers super heavy dark protons.
\end{abstract}


\maketitle
{\it Introduction}
Intensive researches lead to the current standard dark matter,
a paradigm of cold, collisionless particles which has achieved
remarkable success in describing the universe at large scales:  the
cosmic microwave background radiation (CMB) and the large structure of the universe~\cite{Bahcall:1999xn, Springel:2006vs, Trujillo-Gomez:2010jbn}.  However, in the  small scale of the universe such as clusters and galaxies, $\Lambda$CDM does  have some discrepancies or anomalies: ``missing satellites''~\cite{Kauffmann:1993gv, Moore:1999nt}, 
``cusp vs. core''~\cite{Burkert:1995yz,Salucci:2007tm},  ``too big to fail'' 
~\cite{Boylan-Kolchin:2011qkt} and ``diversity''  problems~\cite{Oman:2015xda}.
Although there are several astrophysical solutions to these problems,  these discrepancies require the cross-sections per mass unit $\sigma/m$ of the dark matter at about $1$ cm$^2$/g from the perspective of particle physics, ~\cite{Randall:2008ppe, Robertson:2016xjh}.

In light of this, the self-interacting dark matter (SIDM)  scenario  
can address these anomalies in the small structure
elegantly~\cite{Vogelsberger:2012ku, Rocha:2012jg,
Peter:2012jh,Tulin:2013teo, Wang:2014kja}. The crucial point of the SIDM is that the collision process between the dark matter in the current epoch of our universe is non-relativistic, where the long-range quantum effect dominates in the collision to enhance the cross-section. As a consequence, rich structures are formed at the galaxy scale.  
In addition to the requirement of the large cross-section, 
the properties of the velocity dependence play an important role
in the small structure observations. As the different cross-sections per mass unit are required by the different small scale observations such as
milky way~\cite{Zavala:2012us},  galaxies~\cite{Spethmann:2016glr} 
and clusters~\cite{Kaplinghat:2015aga} with 
different average velocities of the halos.
Roughly speaking, a decreased cross-section in the velocity is required for the SIDM.
(For the detail, see the following context~\cite{Tulin:2017ara, Bullock:2017xww}.)
As the long-range of the quantum effect is much more complicated
than the collisionless particle, the properties of the self-scattering
dark matter is of great interest in recent studies. For example,
the authors proposed a ``puffy'' dark matter in  Ref.~\cite{Chu:2018faw} which can realize the required velocity dependence. In this scenario,
the SIDM with a finite size that is larger than its Compton wave-length
can decrease the cross-sections with the corresponding velocity of the dark matter,
even in the situation of a long-range force.  The finite-size effect can dominate 
in such cases and is independent of the inner structure of the dark matter.
Some QCD-like theories can realize this puffy dark matter and the direct detection
can be relevant to the size of the dark 
matter~\cite{Kusenko:1997si, Kusenko:1997vp,
Kusenko:2001vu,Kitano:2016ooc}.

If the puffy dark matter exists, the scattering between the dark matter will be of great interest as there must be some inner interaction
or structure of the dark matter for which the finite-size must be some residual effect of the intrinsic interaction. For example, 
the Van der Waals force between the molecular is
the residual effect of the interaction between electrons 
and nuclei.~The interactions between the confined quarks 
determine the interaction between
the elastic scattering between hadrons like protons.
We can not ignore the residue effect and consider its implication 
on dark matter self-scattering.
Fortunately, we have enough knowledge of ordinary matters.
It can give us many hints on the dark sector. 
The scattering between finite-size dark matter
is similar to that between protons. The results are
consistent with the  scattering theory that includes both Coulomb and Nuclear interaction
with only very few input parameters~\cite{Block:1984ru}, 
even though the detailed inner structure of the component quarks with strong interactions in the proton are unknown to us.
With this triumph,
 we can apply the same paradigm to dark matter scattering with finite size. In this paper, the Eikonal approximation used 
in $pp$ elastic collision will be adopted and the detailed studies on the scattering
of puffy dark matter are shown in the following. 


{\it The scattering theory of a puffy particle}\label{theory}~~
Before going into the detailed studies on the puffy particle, we must dwell on the general scattering quantum theory. In principle, analyticity and unitarity are fundamental elements to our understanding of particle physics. Analyticity requires that the forward scattering amplitudes for the scattering of the fundamental particles come from analytic functions. Further, the unitarity provides the relation between
the total cross-section and the imaginary portion of the forward scattering amplitude.
Generally, the elastic scattering is the function of the impact parameter that is didactic. We can derive numerous theorems relevant to elastic scattering using this approach. We begin our derivation by following
the standard scattering theory. As discussed about in the introduction, the long-range force
could be the residual effect of the inner interaction or structure of the puffy dark matter, the amplitude is 
\ba
f_c&=&m\frac{\alpha G^{2}(q)}{q^{2}+m_{\phi}^{2}}
\label{eqn:Coulomb}
\ea
for simplicity, in which $m$ is the mass of the scattering particle, 
$\alpha$ is the coupling strength,  
$q^2$ is the transfer momentum,  $m_\phi$ is the mass of the mediator, and $G(q)$ is the form factor. In the case of $m_\phi$ approaching zero, the amplitude
will become a Coulomb scattering, thus a subscript ``$c$'' is denoted.
From the theoretical side, such Coulomb interaction
(or Yukawa interaction in case of a non-zero mass mediator)
can come from some specific U(1) gauge symmetry~\cite{Fabbrichesi:2020wbt}
if it is not ordinary electromagnetic interactions.
The Coulomb scattering from mediators becomes sufficiently sizable only at small $q^2$ 
with a suppressed contribution to the transfer cross-section discussed in the 
following section. 

The amplitude $f_c$ listed above is the Born approximation.
As shown in the 
Refs.~\cite{Tulin:2013teo, Wang:2014kja}, the small scale problem
can be solved in the quantum resonant regime with the angular momentum $l$
at the order of $10$. Nevertheless,
if the particle is puffy or has a finite size i.e. 
$G(q)\neq 1$, additional
interactions exist at a smaller range, and the interaction is strong enough
to confine the U(1) charge in a specific space. As shown in 
Ref.~\cite{Chu:2018faw}, the authors propose
a QCD-like confining theory to realize the puffy dark matter. The form factor $G(q)$ also generates the desirable velocity-dependent behavior of self-interaction dark matter. However, the equation~(\ref{eqn:Coulomb}) does not capture the dark QCD interaction that is responsible for the formation of puffy dark matter. We call it Nuclear interaction to distinguish it from the Coulomb one.

In terms of similarity with the elastic scattering of the protons, 
it suffices to apply the well-developed method to dark protons with Nucleus interaction.  We can develop a geometrical picture-based impact-parameter space. The standard partial-wave expansion of the scattering
amplitude in terms of Legendre polynomials from QCD interaction in the center mass frame is 
\be
f_{n}(s,t)=\frac{1}{k}\sum^{\infty}_{l=0}(2l+1)P_l(\cos\theta)a_l(k),
\ee
Here we use subscript $n$ to stand for Nucleus interaction. 
$s, t$ are the Mandelstam variables,
$k$ is the momentum  of the incident particle and $l$th partial wave is 
\be
a_l(k)=\frac{\exp(2i\delta_l)-1}{2i}.
\ee
$\delta_l$ is the phase shift of $l$th partial wave. 
Note that the subscript ``$n$'' denotes the strong interaction in the 
short range. As the partial wave goes to infinity,
\be
P_l(\cos\theta) \to J_0[(2l+1)\sin(\theta/2)].
\ee
where $J_0(t)$ is the  zero order Bessel function
$J_0(t)=({1}/{2\pi})\int d\phi \exp(-it\cos\phi)$. 
The summation of $l$ can be converted to an integral
\be
f_{n}(s,t)=2k\int b{\rm d}b J_0(qb) a(b,s),
\label{eqn:Nucleus}
\ee
in which the relation between  the newly defined impact parameter  $b$ and  angular momentum   is  $b k = l + 1/2$. 
Then  $a_l \to a(b,s)$ become a function of the impact parameter.
It is perpendicular to the beam direction. We can view it as a distribution
of wave sources that produce an interference pattern. 
We can see that the impact parameter is the 
Fourier transformation of $f_n(s,t)$ in two dimension space, thus
the inversion of $f_n(s,t)$ gives 
\begin{eqnarray}
a(b,s) = \frac{1}{4\pi k}\int {\rm d}^2q\exp(-i{\bm q}\cdot {\bm b})f_n(s,t).
\end{eqnarray}
One can easily check that after removing a $k$ factor, 
$f_n$ only depends on $t$ which 
is always assumed in elastic scattering, then  $a$ will only depend
on impact parameter $b$.
Note that the above methodology is the so-called Eikonal approximation  
which is used for the situation that the wavelength  $\lambda_d$
much less than the size of the collision particles in the 
literature. Thus high-energy
collision or relativistic collision is needed in the $pp$ collision.
For the slowly moving puffy dark matter, 
such approximation would 
be applicable in the case of that the de Broglie wavelength of the heavy dark matter
are much smaller than the size of the dark matter. 
 Together with the appropriate high angular momentum discussed above,
 the application of Eikonal approximation
is adopted here with the following parametrization.
Note that, in the original puffy dark matter paper Ref.~\cite{Chu:2018faw}, the
required suppression behavior $1/v^4$ of the cross section are in the regime 
$mv\gg r_{\rm DM}^{-1}$ which is just the applicable regime for the Eikonal
approximation. 

By above derivation, we can obtain the realistic scattering cross-section including both the Coulomb scattering in Eq.~(\ref{eqn:Coulomb}) and Nucleus scattering
in Eq.~(\ref{eqn:Nucleus}). A phase factor  $\alpha \phi(q)$ 
is introduced to address the interference between Coulomb and strong
interactions. The complete differential cross-section  ~\cite{Xu:2020feo} is 
\be \label{fullcs}
\frac{d\sigma}{d\Omega}=|f_{c}(q)e^{i\alpha\phi(q)}+f_{n}(q)|^{2}
=\frac{d\sigma_{c}}{d\Omega}+\frac{d\sigma_{\rm int}}{d\Omega}+\frac{d\sigma_{n}}{d\Omega},
\ee
where
\ba \label{eq2}
\frac{d\sigma_{c}}{d\Omega}&=&m^{2}\frac{\alpha^{2}G^{4}(q)}{(q^2+m_{\phi}^{2})^{2}},\\
\frac{d\sigma_{\rm int}}{d\Omega}&=&\frac{-m^{2}v}{4\pi}\frac{\alpha\sigma_{\rm tot}G^{2}(q)}{(q^{2}+m_{\phi}^{2})} \nonumber\\  
&&\times e^{-\frac{B|t|}{2}}\left[\rho\cos(\alpha\phi)+\sin(\alpha\phi)\right],\\
\frac{d\sigma_{n}}{d\Omega}&=&
\frac{(mv)^{2}}{4\pi}\frac{\sigma_{\rm tot}^{2}(1+\rho^{2})e^{-B q^2}}{16\pi}.\label{nn}
\ea
Here $v$ is the relative velocity between the scattering particles.
$\phi(q)$ is the  phase that is deeply investigated 
in the literature, and the result from the Eikonal approach  gives
~\cite{Bethe:1958zz,West:1968du,Cahn:1982nr}
\ba
 \phi(q)&=&-\gamma+\ln(\frac{B q^2}{2})+\ln(1+\frac{8 r_0^{2}}{B})\nonumber\\
&& +\ln(4 q^2  r_0^{2})\cdot(4 q^2  r_0^{2})+2 q^2  r_0^{2},
\ea
where $\gamma$ is Euler's constant and $r_0$ is 
the radius of charge distribution of the puffy particle. 
For the Eq. (\ref{nn}), it is originally from  the parametrization of  the nuclear elastic cross section $\frac{d\sigma}{dt} =\left[\frac{d\sigma}{dt}\right]_{t=0}e^{Bt}$.
The factor $e^{Bt}$ is a Gaussian form factor indeed. Note that this parametrization is only an effective approach in 
the Eikonal approximation, the processes  similar to  the 
Deep Inelastic Scattering or Drell-Yan processes in the inelastic scattering of proton
need further explorations. Also, note that $B$ is an adequate parameter
inspired from the ordinary hadron
collision~\cite{Selyugin:2020foq,Selyugin:2021his,Obikhod:2021duk} as
no prior form factor exists for the puffy dark matter.

It is easy to find the Nucleus differential cross-section is the function of $\sigma_{\mathrm{tot}}$, $\rho$ and $B$. We can derive the cross-section from the analyticity and unitarity conditions.  For example, based on the optical theorem
\be
\sigma_{\rm tot} = \frac{4\pi}{k} {\rm Im} f_n(s,0) = 4\int {\rm d}^2 b {\rm Im} a(b,s).
\label{caltot}
\ee
For the detailed study on the scattering impact-parameter space and the application
in $pp$ scattering, one can see Ref.~\cite{Block:1984ru}.  $\sigma_{\rm tot}$ is the total cross-section
which includes all the elastic, inelastic, and annihilation cross-sections.
$\rho= {\rm Re} f_n(0)/{\rm Im} f_n(0)$
is the ratio between real and imaginary part of the scattering amplitude.
$B$ is the slope parameter when we expand the cross section with $t$
\be
B(s,t) = \frac{{\rm d}}{{\rm d}t }\left[\ln \frac{{\rm d}\sigma}{{\rm d}t}\right].
\ee 
$B(s,t)$ measures the size of the puffy particle.
In the case of $pp$ scattering, we can extract these parameters from the experiments
in Coulomb normalization or the ``luminosity-free'' method.
Note that there is a fourth curvature parameter $C$ when we expand $t$ to the
subleading order, which we neglect in this work as it is the higher-order result. From the theoretical side, all the three parameters and other observable should be predicted or calculated
from impact-parameter space. As an example, when $a(b,s)$ is purely imaginary, the 
elastic scattering cross-section would be~\cite{Block:1984ru}
\be
\sigma_{\rm el} = 4\int {\rm d}^2b \left({\rm Im} a\right)^2,
\label{calel}
\ee
and $B$ can be derived from 
\be
\sigma_{\rm tot} B = 2\int {\rm d}^2b b^2  {\rm Im} a.\label{calB}
\ee
Thus we can use the amplitude $a(b,s)$ to derive the scattering
cross-section by 
using Eq.(\ref{caltot}), Eq.(\ref{calel}) and Eq.(\ref{calB}).

In the above, we show the sketch map of the quantum scattering 
theory of a puffy particle.
We can see that all the concerns lay on the impact-parameter amplitude 
$a(b,s)$.  As shown in the Ref.~\cite{Block:1984ru}, pure
imaginary $a(b,s)$ are always adopted for the forward
high-energy scattering, there are several profiles
such as disk, parabolic form, Gaussian  shape and Chou-Yang model {\it etc. }
which can be used in the literature.
Naively thinking, the puffy particle shows a 
$U(1)$ charge locates in finite-size space.  
Chou-Yang model  proposed in Ref.~\cite{Chou:1968bc, Chou:1983zi,
Durand:1968ny}  postulates that
the charge distribution should have a similar profile or be determined by the impact-parameter amplitude. The attenuation of the amplitude accompanying
the process of the two puffy particles going through each other is governed 
by the local opaqueness within each particle. It means that the charge distribution or the electro-magnetic form factor should
have the same shape as the transverse distribution of the matter,
leaving only the strength of the absorption to be fixed. In the visible sector, we require the total cross-section calculated 
in the model to agree with the experiments.
In the dark sector, three more factors are introduced to account for the complete scattering process. From the point of view of minimality and predictivity, this more realistic model, the Chou-Yang model, 
can perfectly solve the problem.
In the model, the absorption at an impact parameter $b$ is $\Omega(b)$, then
\be\label{aa}
a(b,s) =\frac{\exp(2i\delta)-1}{2i}\equiv \frac{i}{2}\left[1-\exp(-\Omega(b))\right].
\ee
If the electro-magnetic form factor is dipole
\be
G(q) =\left(\frac{1}{1+r_0^2 q^2}\right)^2,
\ee
using the Fourier transform of 
the matter distribution, then a convolution of the dipole form factor shows
that $\Omega$ is analytically
\be
\Omega = \frac{A}{8}x^3K_3(x),
\ee
in which $x=b/r_0$, $A$ is a dimensionless parameter
which we call amplitude coefficient and 
$K_3$ is the modified Bessel function.
By choosing a suitable parameter $A$, the resulting cross-section agrees with the experiment results. 
It indicates that we can use a similar picture on the puffy dark matter. 
Note that the pure imaginary impact factor $a(b,s)$ 
is generally used in high-energy scattering \cite{Block:1984ru}, 
the validity of such assumption 
for the $t=0$ low energy scattering of puffy dark matter
needs further check. The Eikonal approximation in such case
is discussed in the following section.

{\it Puffy dark matter and the implication on the small cosmological scale}~
We generally require a  large self-scattering cross-section for the small-scale anomalies. However, the formation of the structure is in fact 
dependent on the transfer cross-section  via combining the Eq. (\ref{fullcs}, \ref{caltot}, \ref{calB},\ref{aa})
\be
\sigma_T=\int d\Omega (1- \cos\theta )\frac{{\rm d}\sigma} {{\rm d}\Omega}.
\ee
A significant issue in the solution of the anomalies is the behavior of the velocity dependence, where the radius of the dark matter is substantially larger than the range of the Coulomb force for puffy DM.

\begin{figure}[htpb]
\begin{center}
\vspace{-0.5cm}
\scalebox{0.28}{\hspace{-2.cm}\epsfig{file=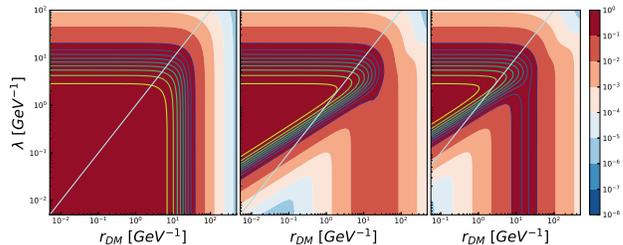}}
\caption{Ratio $\sigma_T/\sigma_A$ in $r_{\rm DM}$  and $\lambda$
space with different absorption  parameter $A=0, 1, 10$, respectively.
 The other parameters are the same which is  $m_{\rm DM}=100$ GeV, 
$\alpha=0.01$ and $v/c=10^{-3}$.}\label{fig1}
\end{center}
\vspace{-0.5cm}
\end{figure}

As shown in Ref.~\cite{Chu:2018faw},
if the interaction is pure dark $U(1)$ interaction, 
the transfer cross-section $\sigma_T$ will be of  $1/v^4$ dependence at $mv  r_{\rm DM} \gg 1$.  Note that
the decrease with the velocity of $\sigma_T$ talked about is on
\be
\sigma_0 = 4\pi(m_{\rm DM} \alpha \lambda^2)^2
\ee 
which is the naively dimensional estimation on the corresponding $U(1)$
cross-section at high energy. The corresponding results are shown
in the left panel of Fig. \ref{fig1}  in which the vertical coordinate 
$\lambda =1/m_\phi$ is the range of $U(1)$ interaction, the horizontal coordinate
$r_{\rm DM}=2 \sqrt{6}r_0$ is the radius of the puffy dark matter.
The upper left part of the left plot 
 gives the correct velocity dependence coinciding with that of the SIDM
with a point particle~\cite{Tulin:2013teo}.
The lower right part shows that ratio $\sigma_T/\sigma_0$
also depends on the radius of the dark matter. Furthermore, 
in the case of a much smaller $\lambda$ that the long-range Coulomb force disappears, the finite-size effect
can still maintain the correct velocity dependence.

However, a flaw or an inconsiderate point for the above results is that we can not ignore the binding force of puffy dark matter.  The scattering theory in Ref~\cite{Block:1984ru} gives us a clue to these questions.
At first, the cross-section at high energy is
an electromagnetic part ($\sigma_0$) plus a strong interaction part,  
\be
\sigma_A = \sigma_0 + \sigma_n,
\ee
To make the impact factor more realistic, we use the Chou-Yang model. Then according to Eq. (\ref{caltot}),  $\sigma_n$ can be defined as
\be
\sigma_n = 8\pi r_{\rm DM}^2 \int {\rm d} x x {\rm Im}a(b,s) .
\ee
It is the total cross-section of the strong interaction.
By this definition, the dipole electromagnetic form factor has the same profile as the matter distribution, and $A$ is the amplitude coefficient.  
Eq. (\ref{fullcs})  and the corresponding calculations can be used for the differential cross-section used for the transfer cross-section. Compared with only the $U(1)$ interaction, we can see that the calculation here would be more realistic. Nevertheless, only one additional
parameter $A$ with definite physical implications could be more predictive. 

We show the ratio of $\sigma_T/\sigma_A$ numerically with
different $A$ in Fig. \ref{fig1}
in which $A=0$ case talked above lists in the left panel.
From Fig.  \ref{fig1} we can see that after the consideration of the strong
interaction, the decrease of the ratio changes 
a lot in the $\lambda$ and $r_{\rm DM}$
space. Total space consists of two parts: the upper left semi-ellipse
zone indicates the $U(1)$ interaction dominants;
the lower right ladder-like zone is the radius effect dominant part. 
Both parts depend on the ratio of radius and force range in the $A=0$ case. 
We can also see that in the case of a larger $A$, the effect of the radius
becomes more significant. 
It can affect the ratio even when the force range $\lambda$
is greater than the radius $r_{\rm DM}$. 

\begin{figure}[htpb]
\begin{center}
\vspace{0.5cm}
\epsfig{file=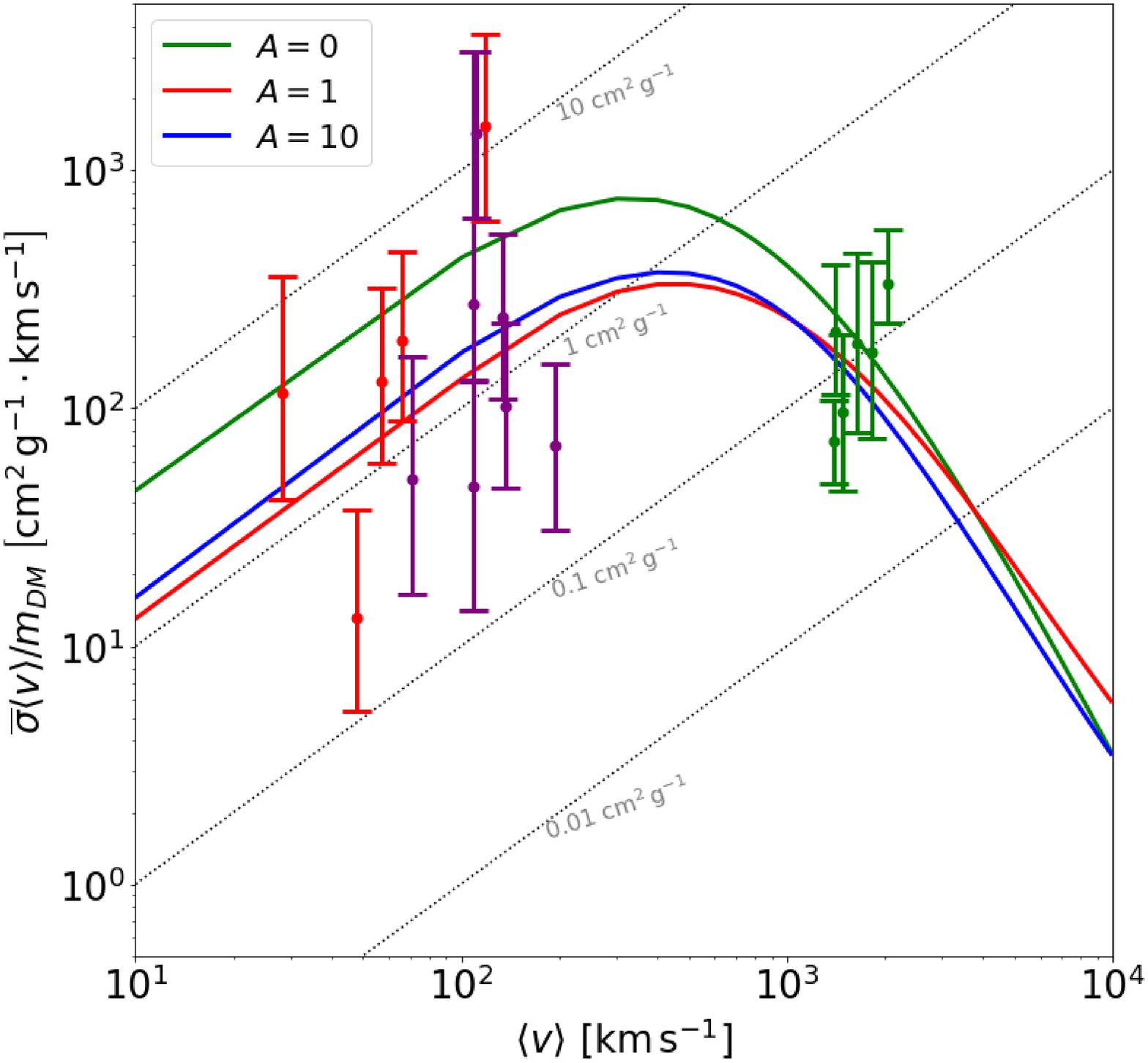,height=4cm,width=4cm}
\epsfig{file=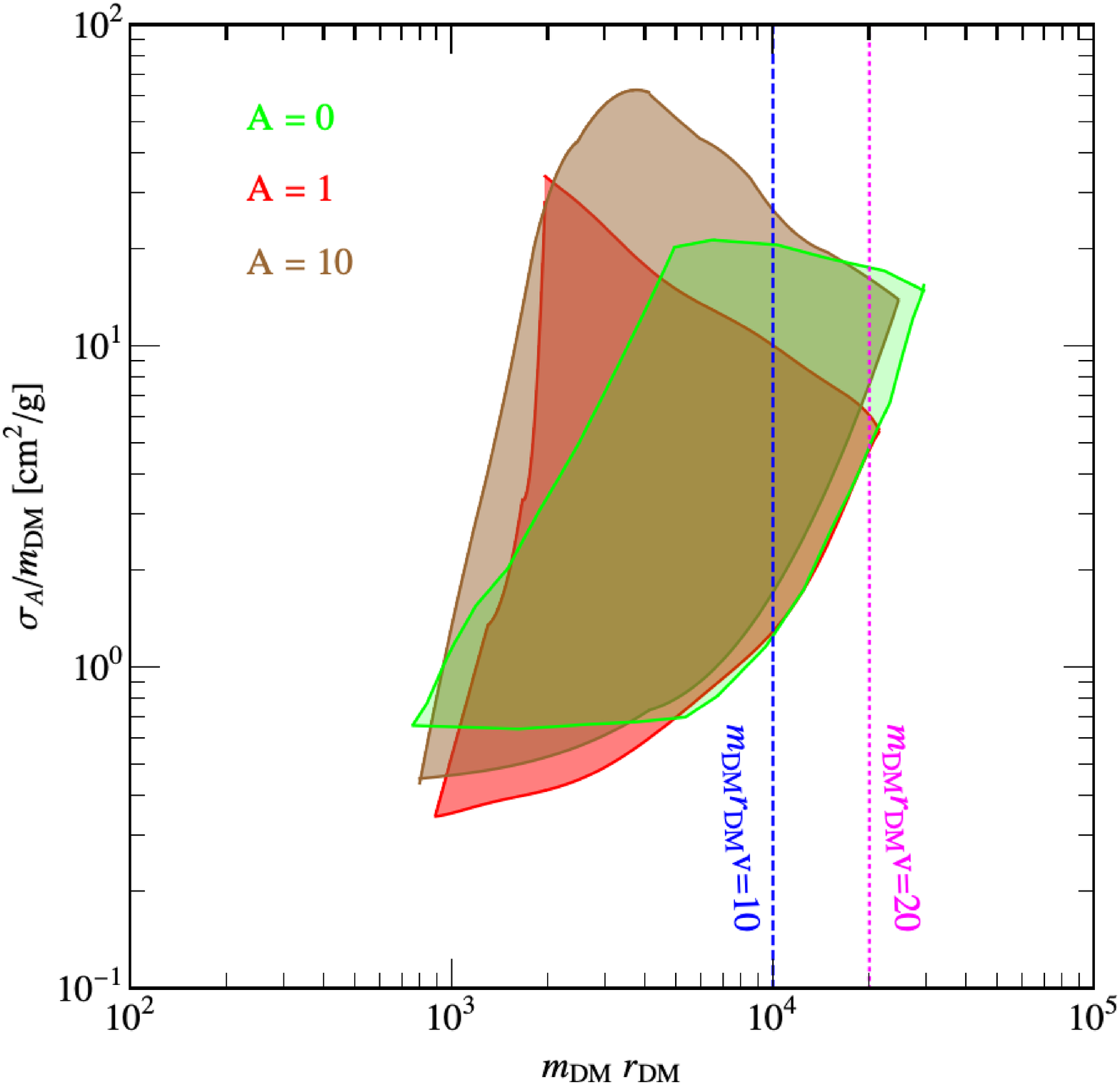,height=4cm,width=4cm}
\caption{Left panel: The best fit curve to data \cite{Kaplinghat:2015aga}  for the velocity dependence 
of the  transfer cross-section of the puffy dark matter in different $A$. 
The observable  red, purple and green data points respectively are from five clusters \cite{Newman:2012nw}, seven low-surface-brightness spiral galaxies \cite{KuziodeNaray:2007qi} and six dwarf galaxies \cite{Oh:2010ea}.
The Corresponding parameters for the curves 
are ($A=0,~m_{\phi}=0.03 \rm {GeV},~r_{\rm DM}=37.19~\rm{GeV^{-1}},~ m_{\rm {DM}}=20.28~\rm{GeV},~\alpha=0.0083$),
($A=1,~m_{\phi}=0.026~\rm {GeV},~r_{\rm DM}=46.52~\rm{GeV^{-1}},~m_{\rm {DM}}=12.47~\rm{GeV},~\alpha=0.004$),
($A=10,~m_{\phi}=0.063~\rm {GeV},~r_{\rm DM}=28.36~\rm{GeV^{-1}},~m_{\rm {DM}}=29.28~\rm{GeV},~\alpha=0.017$).
Right panel: the fitted $\sigma_A/m_{\rm DM}$ 
versus $m_{\rm DM} r_{\rm DM}$ in the 
$\lambda < r_{\rm DM}$ case. The blue dashed line denotes the validity condition for the  Eikonal approximation 
when $v/c=10^{-3}$. }\label{fig2}
\end{center}
\vspace{-0.5cm}
\end{figure}
Next, we calculate the velocity-averaged transfer cross-section in the
Maxwell-Boltzmann distribution to check the constraints in the small
cosmological scale. The case of the best-fit point is in the left panel
of Fig. \ref{fig2} in the different value of $A$.  
The space 95\% C.L.  contours of $\sigma_A/m_{\rm DM}$ 
and $m_{\rm DM}r_{\rm DM}$ in the 
$\lambda < r_{\rm DM}$ case are shown in the right panel 
of the Fig. \ref{fig2}. 
The blue dash line of the right panel shows the edge of   
 the validity of the Eikonal approximation. 
 As mentioned in the previous section, such estimates are valid when the de Broglie
 wavelength is significantly smaller than the size of the puffy dark matter. 
 Here the
condition should be  
\begin{eqnarray}
m_{\rm DM} r_{\rm DM} v\gg 1.
\end{eqnarray}
The blue (purple) dash line shows $m_{\rm DM} r_{\rm DM}v=10$ ($=20$), implying
that we keep the results in about $10\%$ precision limits.
Of course, the right part of the rest excluded region will be more
concrete along with the increasing of radius $r_{\rm DM}$.
Compared with the $A=0$ case,
we can see that the strong interaction will enlarge the range of 
$\sigma_A/m_{\rm DM}$. 
Some space that is not satisfied in the pure $U(1)$ case can be 
available when we considered the inner interactions.
The cross-section can be extended to 3 times larger in the case of $A=10$,
and even larger in other amplitude coefficients.
It means that the effect of the radius is
significant in the exploration of dark matter simulations. In some
spaces, it will dominate the physics of dark matter. Finally, the dark proton can generate the correct relic density by the annihilation cross-section~\cite{Huo:2015nwa}
\begin{equation}
\sigma v_{\mathrm{rel}}^{\mathrm{DM}}\left(m_{\mathrm{DM}}\right)=\sigma v_{\mathrm{rel}}^{\mathrm{QCD}} \times\left(\frac{1 \mathrm{GeV}}{m_{\mathrm{DM}}}\right)^{2}
\end{equation}
where $\sigma v_{\mathrm{rel}}^{\mathrm{QCD}}$ is the proton-anti-proton annihilation cross-section in Standard QCD, which comes from the direct measurement in experiment~\cite{OBELIX:1996pze}. After performing the thermal average, the requirement of $\langle\sigma_{\mathrm{rel}}^{\mathrm{DM}}\rangle=3\times 10^{-26}\mathrm{cm}^3/\mathrm{s}$ indicates the rough estimate of dark proton mass is around $150~\mathrm{TeV}$. However, Unitarity's upper bound on dark matter elastic scattering prevents it from achieving such a large self-scattering cross-section for $m_{\mathrm{DM}} = 150$ TeV. That strongly suggests looking for an additional production mechanism for dark protons to reduce their mass scale.

{\it Conclusion}
If the dark matter is not a point particle but some composite ones located in a finite size space, the scattering between dark matter will be much more complicated.  The Eikonal approximation and the similar 
parametrization as $pp$ collision are adopted. 
The sketch map of the calculation of the cross-section and a more 
realistic realization of the matter and  charge distribution, as used by the dipole form factor in the Chou-Yang model, 
are shown in this work. Simultaneously the Chou-Yang model is also introduced to reduce the number of input
parameters to one based on the simplicity and analyticity principle. We should emphasize that though the absorption approaches
zero the scattering can restore the pure $U(1)$ case, 
the inner interaction and the structure
can not be ignored if the dark matter is puffy from the physical insight. 
The method we found is a more appropriate way to 
our understanding of the nature of the puffy dark matter, 
and we proposed a new definition of velocity dependence in such cases.

The numerical results show that even in the range of sizable $U(1)$ interaction, 
the non-vanishing amplitude coefficient $A$  can also affect the scattering
cross-section of the puffy dark matter.  We find that 
with the participation of the strong interaction,
the space of  the cross-section to the mass ratio  which is needed 
in the simulation can be enlarged, 
giving us a more flexible parameter space to other 
processes related to dark matter.

Though the scattering of the puffy dark matter is realized in the paper,
we should note that 
the main shortcoming of this treatment is that the Eikonal approximation only covers part
of parameters space, which might lose the accuracy in low-velocity dark protons.  The detailed research on the elastic scattering
will give us further insight into the formation of the finite size, such as the QCD-like model, which will be explored in our future work.

\section*{Acknowledgments}
This work was supported by the Natural Science Foundation of China under grant number 11775012 and 11805161. The work of BZ is also supported partially by Korea Research Fellowship Program through the National Research Foundation of Korea (NRF) funded by the Ministry of Science and ICT (2019H1D3A1A01070937).


\begin{thebibliography}{99}
\bibitem{Bahcall:1999xn}
N.~A.~Bahcall, J.~P.~Ostriker, S.~Perlmutter and P.~J.~Steinhardt,
Science \textbf{284} (1999), 1481-1488

\bibitem{Springel:2006vs}
V.~Springel, C.~S.~Frenk and S.~D.~M.~White,
Nature \textbf{440} (2006), 1137

\bibitem{Trujillo-Gomez:2010jbn}
S.~Trujillo-Gomez, A.~Klypin, J.~Primack and A.~J.~Romanowsky,
Astrophys. J. \textbf{742} (2011), 16

\bibitem{Kauffmann:1993gv}
G.~Kauffmann, S.~D.~M.~White and B.~Guiderdoni,
Mon. Not. Roy. Astron. Soc. \textbf{264} (1993), 201

\bibitem{Moore:1999nt}
B.~Moore, S.~Ghigna, F.~Governato, G.~Lake, T.~R.~Quinn, J.~Stadel and P.~Tozzi,
Astrophys. J. Lett. \textbf{524} (1999), L19-L22

\bibitem{Burkert:1995yz}
A.~Burkert,
Astrophys. J. Lett. \textbf{447} (1995), L25

\bibitem{Salucci:2007tm}
P.~Salucci, A.~Lapi, C.~Tonini, G.~Gentile, I.~Yegorova and U.~Klein,
Mon. Not. Roy. Astron. Soc. \textbf{378} (2007), 41-47

\bibitem{Boylan-Kolchin:2011qkt}
M.~Boylan-Kolchin, J.~S.~Bullock and M.~Kaplinghat,
Mon. Not. Roy. Astron. Soc. \textbf{415} (2011), L40

\bibitem{Oman:2015xda}
K.~A.~Oman, J.~F.~Navarro, A.~Fattahi, C.~S.~Frenk, T.~Sawala, S.~D.~M.~White, R.~Bower, R.~A.~Crain, M.~Furlong and M.~Schaller, \textit{et al.}
Mon. Not. Roy. Astron. Soc. \textbf{452} (2015) no.4, 3650-3665

\bibitem{Randall:2008ppe}
S.~W.~Randall, M.~Markevitch, D.~Clowe, A.~H.~Gonzalez and M.~Bradac,
Astrophys. J. \textbf{679} (2008), 1173-1180

\bibitem{Robertson:2016xjh}
A.~Robertson, R.~Massey and V.~Eke,
Mon. Not. Roy. Astron. Soc. \textbf{465} (2017) no.1, 569-587

\bibitem{Vogelsberger:2012ku}
M.~Vogelsberger, J.~Zavala and A.~Loeb,
Mon. Not. Roy. Astron. Soc. \textbf{423} (2012), 3740

\bibitem{Rocha:2012jg}
M.~Rocha, A.~H.~G.~Peter, J.~S.~Bullock, M.~Kaplinghat, S.~Garrison-Kimmel, J.~Onorbe and L.~A.~Moustakas,
Mon. Not. Roy. Astron. Soc. \textbf{430} (2013), 81-104

\bibitem{Peter:2012jh}
A.~H.~G.~Peter, M.~Rocha, J.~S.~Bullock and M.~Kaplinghat,
Mon. Not. Roy. Astron. Soc. \textbf{430} (2013), 105

\bibitem{Tulin:2013teo}
S.~Tulin, H.~B.~Yu and K.~M.~Zurek,
Phys. Rev. D \textbf{87} (2013) no.11, 115007

\bibitem{Wang:2014kja}
F.~Wang, W.~Wang, J.~M.~Yang and S.~Zhou,
Phys. Rev. D \textbf{90} (2014) no.3, 035028

\bibitem{Zavala:2012us}
J.~Zavala, M.~Vogelsberger and M.~G.~Walker,
Mon. Not. Roy. Astron. Soc. \textbf{431} (2013), L20-L24

\bibitem{Spethmann:2016glr}
C.~Spethmann, H.~Veerm\"ae, T.~Sepp, M.~Heikinheimo, B.~Deshev, A.~Hektor and M.~Raidal,
Astron. Astrophys. \textbf{608} (2017), A125

\bibitem{Kaplinghat:2015aga}
M.~Kaplinghat, S.~Tulin and H.~B.~Yu,
Phys. Rev. Lett. \textbf{116} (2016) no.4, 041302

\bibitem{Tulin:2017ara}
S.~Tulin and H.~B.~Yu,
Phys. Rept. \textbf{730} (2018), 1-57

\bibitem{Bullock:2017xww}
J.~S.~Bullock and M.~Boylan-Kolchin,
Ann. Rev. Astron. Astrophys. \textbf{55} (2017), 343-387

\bibitem{Chu:2018faw}
X.~Chu, C.~Garcia-Cely and H.~Murayama,
Phys.\ Rev.\ Lett.\  {\bf 124}, no. 4, 041101 (2020)

\bibitem{Kusenko:1997si}
A.~Kusenko and M.~E.~Shaposhnikov,
Phys. Lett. B \textbf{418} (1998), 46-54

\bibitem{Kusenko:1997vp}
A.~Kusenko, V.~Kuzmin, M.~E.~Shaposhnikov and P.~G.~Tinyakov,
Phys. Rev. Lett. \textbf{80} (1998), 3185-3188

\bibitem{Kusenko:2001vu}
A.~Kusenko and P.~J.~Steinhardt,
Phys. Rev. Lett. \textbf{87} (2001), 141301

\bibitem{Kitano:2016ooc}
R.~Kitano and M.~Kurachi,
JHEP \textbf{07} (2016), 037


\bibitem{Block:1984ru}
M.~M.~Block and R.~N.~Cahn,
Rev.\ Mod.\ Phys.\  {\bf 57}, 563 (1985).

\bibitem{Fabbrichesi:2020wbt}
M.~Fabbrichesi, E.~Gabrielli and G.~Lanfranchi,
[arXiv:2005.01515 [hep-ph]].

\bibitem{Xu:2020feo}
H.~Xu, Y.~Zhou, U.~Bechstedt, J.~B\"oker, A.~Gillitzer, F.~Goldenbaum, D.~Grzonka, Q.~Hu, A.~Khoukaz and F.~Klehr, \textit{et al.}
Phys. Lett. B \textbf{812} (2021), 136022

\bibitem{Selyugin:2020foq}
O.~V.~Selyugin,
Symmetry \textbf{13} (2021), 164

\bibitem{Selyugin:2021his}
O.~V.~Selyugin,
Phys. Rev. D \textbf{104} (2021) no.3, 034001
doi:10.1103/PhysRevD.104.034001
[arXiv:2107.02514 [hep-ph]].

\bibitem{Obikhod:2021duk}
T.~Obikhod and I.~Petrenko,
[arXiv:2107.07411 [hep-ph]].

\bibitem{Bethe:1958zz}
H.~A.~Bethe,
Annals Phys. \textbf{3} (1958), 190-240

\bibitem{West:1968du}
G.~B.~West and D.~R.~Yennie,
Phys. Rev. \textbf{172} (1968), 1413-1422

\bibitem{Cahn:1982nr}
R.~Cahn,
Z. Phys. C \textbf{15} (1982), 253

\bibitem{Chou:1968bc}
T.~T.~Chou and C.~N.~Yang,
Phys. Rev. \textbf{170} (1968), 1591-1596

\bibitem{Chou:1983zi}
T.~T.~Chou and C.~N.~Yang,
Phys. Lett. B \textbf{128} (1983), 457-460


\bibitem{Durand:1968ny}
L.~Durand, III and R.~Lipes,
Phys. Rev. Lett. \textbf{20} (1968), 637-640
doi:10.1103/PhysRevLett.20.637

\bibitem{Kaplinghat:2015aga}
M.~Kaplinghat, S.~Tulin and H.~B.~Yu,
Phys. Rev. Lett. \textbf{116} (2016) no.4, 041302
doi:10.1103/PhysRevLett.116.041302
[arXiv:1508.03339 [astro-ph.CO]].

\bibitem{Newman:2012nw}
A.~B.~Newman, T.~Treu, R.~S.~Ellis and D.~J.~Sand,
Astrophys. J. \textbf{765} (2013), 25
doi:10.1088/0004-637X/765/1/25
[arXiv:1209.1392 [astro-ph.CO]].

\bibitem{KuziodeNaray:2007qi}
R.~Kuzio de Naray, S.~S.~McGaugh and W.~J.~G.~de Blok,
Astrophys. J. \textbf{676} (2008), 920-943
doi:10.1086/527543
[arXiv:0712.0860 [astro-ph]].

\bibitem{Oh:2010ea}
S.~H.~Oh, W.~J.~G.~de Blok, E.~Brinks, F.~Walter and R.~C.~Kennicutt, Jr,
Astron. J. \textbf{141} (2011), 193
doi:10.1088/0004-6256/141/6/193
[arXiv:1011.0899 [astro-ph.CO]].

\bibitem{Huo:2015nwa}
R.~Huo, S.~Matsumoto, Y.~L.~Sming Tsai and T.~T.~Yanagida,
JHEP \textbf{09} (2016), 162

\bibitem{OBELIX:1996pze}
A.~Bertin \textit{et al.} [OBELIX],
Phys. Lett. B \textbf{369} (1996), 77-85



\end{thebibliography}
\end{document}